
\input harvmac

\Title{\vbox{\baselineskip12pt\hbox{RIMS-878}}}
{\vbox{\centerline{Topological Lattice Models in Four Dimensions}}}

\centerline{Hirosi OOGURI \foot{e-mail: ooguri@kekvax.kek.ac.jp,
ooguri@jpnyitp.bitnet}}
\vskip .3cm
\centerline{Research Institute for Mathematical Sciences}
\centerline{Kyoto University, Sakyo-ku, Kyoto 606-01, Japan}

\vskip 1.5cm
\centerline{{\it Dedicated to Professors Huzihiro Araki
and Noboru Nakanishi}}
\centerline{{\it on the occasion of their sixtieth birthdays}}

\vskip 1.1 cm
We define a lattice statistical model on a triangulated
manifold in
four dimensions associated to a group $G$.
When $G=SU(2)$, the statistical weight is constructed
from the $15j$-symbol as well as the $6j$-symbol for
recombination of angular momenta, and the model may be
regarded as the four-dimensional version of the Ponzano-Regge
model. We show that the partition function of the model
is invariant under the Alexander moves of the simplicial complex,
thus it depends only on the piecewise linear topology of
the manifold. For an orientable manifold, the model is
related to the so-called $BF$ model. The $q$-analogue
of the model is also constructed,
and it is argued that its partition function is
invariant under the Alexander moves.
It is discussed
how to realize the 't Hooft operator in these models
associated to a closed surface in four dimensions
as well as the Wilson operator associated
to a closed loop. Correlation functions of
these operators in the $q$-deformed version of the model
would define a new type of invariants of knots and links
in four dimensions.

\Date{May, 1992}


Recently
several numerical results have been reported
in the three and four-dimensional simplicial quantum gravities
concerning their phase structure \ref\three{M.E. Agishtein and
A.A. Migdal, {\sl Mod. Phys. Lett.} {\bf A6} (1991)
1863; J. Ambjorn and S. Varsted, {\sl Phys. Lett.}
{\bf B226} (1991) 258; D.V. Boulatov and
A. Krzywicki, {\sl Mod. Phys. Letts} {\bf A6}
(1991) 3005; J. Ambjorn, D. Boulatov, A. Krzywicki and
S. Varsted, preprint NBI-HE-91-46 (1991).},
\ref\four{M.E. Agishtein and A.A. Migdal, preprint
PUPT-1287 (1991), PUPT- (1992); J. Ambjorn and J. Jurkiewicz,
preprint NBI-HE-91-47 (1991).}.
Especially in four dimensions, it is suggested in \four\ that
there is a second order phase transition point
indicating the continuum limit of the model.
In view of the importance of this suggestion, it is
desirable to develop an alternative analytical approach
to the dynamical triangulations.

In two dimensions, the matrix model \ref\two{E. Brezin, C. Itzykson,
G. Parisi and J.-B. Zuber, {\sl Commun. Math. Phys.}
{\bf 59} (1978) 35.} has been found a useful tool
in studying the simplicial gravity,
and there have been some attempts to extend it to higher
dimensions \ref\tensor{J. Ambjorn, B. Durhuus and T. Jonsson,
{\sl Mod. Phys. Lett.} {\bf A6} (1991) 1133; N. Sasakura,
{\sl Mod. Phys. Lett.} {\bf A6} (1991) 2613;
M. Douglas and H. Ooguri, unpublished;
N. Godfrey and M. Gross, {\sl Phys. Rev.} {\bf D43}
(1991) R1749.}. In this so-called tensor model,
one considers an integral over
rank-$3$ tensors rather than matrices,
and the perturbative expansion of
the integral generates randomly triangulated three-dimensional
manifolds. However, in this model, it is found difficult
to have control over topologies generated by the random triangulation.
This is because
the model does not contain the sufficient number of parameters
to distinguish different topologies.
To solve this problem,
Boulatov \ref\boulatov{D.V. Boulatov, preprint SPhT/92-017
(1992)} proposed an improved version of the model
defined in terms of functions with three variables on a
group or a quantum group rather than the tensors.
In his model, the perturbative expansion of
the integral is expressed as a sum over triangulated three-manifolds
weighted with some topological invariant
which depends on the choice of the group $G$. There is a
large variety of topological weights corresponding to
various choices of groups, and
one may hope to gain control in summing over topologies
with suitable choice of groups.

The purpose of this paper is two-folded.
First we will extend the Boulatov's model to four dimensions and
check that its perturbative
expansion is expressed as a sum over triangulated four-manifolds.
We then show that the weight for the summation is invariant under
the Alexander moves \ref\alex{J.W. Alexander, {\sl Ann. Math.}
{\bf 31} (1930) 292.} \ref\move{M. Gross and S. Varsted,
preprint NBI-HE-91-33 (1991)} of the simplicial complex, and thus depends
only on the combinatorially equivalent class of the triangulation.
This model could be useful for analytical study of the four-dimensional
simplicial gravity. It is known that
there is no algorithmical way to classify
topologies in four dimensions, but
it would still be possible that, for a suitable choice of $G$,
the perturbative expansion of the model is dominated by
$S^4$ or by some restricted set of topologies
\foot{This seems to be the
case in three dimensions \boulatov .}.

The topological invariants obtained in this way
are interesting in their own right, and
examining their properties is the second purpose of this
paper \foot{In this paper, we discuss geometries in
the piecewise linear category.
If two differential manifolds are diffeomorphic,
their smooth triangulations are combinatorially equivalent
\ref\haupt{J.R Munkres, {\it Elementary Differential Topology,
Annals of Mathematics Studies 54},
(Princeton University Press, 1963).}.
Thus an invariant of the piecewise linear topology gives
a diffeomorphism invariant of smooth manifolds.
In four dimensions,
it is not know whether a homeomorphism of two
combinatorial manifolds also
implies their combinatorial equivalence.
This is known to be the case in two and three dimensions.}.
In order to solve the quantum gravity in four dimensions,
independently of how one defines it,
one would need to have a good understanding
on four-dimensional geometries.
Thus it should be useful to study
various geometrical invariants formulated in the
language of quantum field theory and statistical mechanics.
In the three-dimensional model of Boulatov,
if we choose $G$ to be the quantum group $SU_q(2)$,
the weight in summing over topologies
is constructed from the $6j$-symbols of $SU_q(2)$
\ref\kir{A.N. Kirillov and
N.Yu. Reshetikhin, in {\it Infinite Dimensional
Lie Algebras and Groups}, edited by V.G. Kac
(World Scientific, 1989).}.
The weight then
reproduces the three-dimensional topological invariant
defined by Turaev and Viro \ref\viro{V.G.
Turaev and O.Y. Viro, ``{\it State Sum Invariants of
$3$-Manifolds and Quantum $6j$-Symbols},''
preprint (1990)}, whose $q \rightarrow 1$
limit has been studied many years ago
by Ponzano and Regge \ref\ponzano{G. Ponzano and T. Regge,
in {\it Spectroscopic and Group Theoretical Methods in
Physics}, ed. F. Bloch (North-Holland, Amsterdam, 1968).}.
It has been shown in \ref\turaev{V.G. Turaev, {\sl C. R. Acad. Sci. Paris,
t. 313, S\'erie I} (1991) 395; K. Walker, ``{\it  On Witten's
$3$-Manifold Invariants},''
preprint (1990).}
\ref\ooguri{H. Ooguri and N. Sasakura, {\sl Mod. Phys. Lett.}
{\bf A6} (1991) 3591; H. Ooguri, preprint RIMS-851 (1991),
to be published in {\sl Nucl. Phys. B}.}
that the Turaev-Viro invariant
is related to the partition function of the Chern-Simons gauge theory
\ref\witten{E. Witten, {\sl Commun. Math. Phys.}
{\bf 121} (1989) 351; {\sl Nucl. Phys.} {\bf B311}
(1988/89) 46.} \ref\resh{N. Reshetikhin and V.G. Turaev,
{\sl Invent. Math.} {\bf 103} (1991) 547.}
in three dimensions. We shall see that
this feature persists in four dimensions.
For $q=1$, the four-dimensional model
constructed here makes use of the $15j$-symbols as well
as the $6j$-symbols, and it may be regarded as a
four-dimensional version of the Ponzano-Regge model.
It is also possible define the $q$-analogue of this model,
whose partition function could be regarded as a four-dimensional
version of the Turaev-Viro invariant.
Although the proof of the topological invariance given here
only applies to the case of $q=1$, we present an evidence
that it should also hold for when $q$ is a root of unity.
For $q=1$, the lattice model is related to
the so-called $BF$ model \ref\schwarz{A.S. Schwarz,
{\sl Lett. Math. Phys.} {\bf 2} (1978) 247; {\sl
Commun. Math. Phys.} {\bf 67} (1979) 1.} \ref\horowitz{
G. Horowitz, {\sl Commun. Math. Phys.} {\bf 125} (1989)
417.} \ref\blau{M. Blau and G. Thompson,
{\sl Ann. Phys.} {\bf 209} (1991) 129.}.
We have not been able to identify
a quantum field theory corresponding to the $q$-analogue model,
and this task is left for future investigation.

In the three-dimensional Chern-Simons theory,
the Wilson operators \ref\wilson{K.G. Wilson,
{\sl Phys. Rev.} {\bf D16} (1974) 2445. } along closed loops
make a basic set of observables
and their expectation values are related to the Jones
invariants of knots \witten . One may also consider
the 't Hooft operator \ref\thooft{G. 't Hooft, {\sl
Nucl. Phys.} {\bf B138} (1978) 1.}, but
it is known that the 't Hooft operator is identical to
the Wilson operator in the Chern-Simons theory \ref\moore{G. Moore
and N. Seiberg, {\sl Phys. Lett} {\bf B220} (1989) 422.}.
In four dimensions, on the other
hand, it is a surface that makes a non-trivial knot.
We study the 't Hooft operator in the $BF$ model,
which is associated to $S^2$ embedded in four dimensions and
differs from the Wilson operator, and discuss how to
realize it in the lattice model.
Both the 't Hooft operator and the Wilson operator
create {\it physical states} on $S^2 \times S^1$ embedded in the
four-dimensional manifold $M$. In the case of the 't Hooft operator,
$S^2 \times S^1$ is a boundary of $S^2 \times D^2$ embedded in $M$,
and its $S^1$-part is contractable.
On the other hand, for the Wilson operator,
$S^2 \times S^1$ is a boundary of $D^3 \times S^1$, and
the $S^1$-part may link with other operators.
Correlation functions of the these operators
in the $q$-deformed version of our lattice model
would define a new type of invariants of knots and links
in four dimensions.

Let us define our model. We start with a real-valued function
of four variables $\phi(g_1,g_2,g_3,g_4)$ on $G$ ($g_i \in G$).
For simplicity of notations, we describe the case when
$G=SU(2)$, but extensions to other groups should be obvious.
If we take $G$ to be a finite group, the model constructed below
may be regarded as a four-dimensional version of the
Dijkgraaf-Witten model \ref\dijk{R. Dijkgraaf and E. Witten,
{\sl Commun. Math. Phys.} {\bf 129} (1990) 393.}.
The function $\phi$ can be expanded in terms of matrix elements
$D^j_{mn}(g)$ of a spin-$j$ representation as
\eqn\general{ \phi(g_1,g_2,g_3,g_4)
    = \sum_{j_i,m_i,n_i}
     \phi^{j_1...j_4}_{m_1 n_1...m_4 n_4}
      D^{j_1}_{m_1 n_1}(g_1) \cdots D^{j_4}_{m_4 n_4}(g_4).}
We then require $\phi$ to be invariant under the right action
of $G$,
$$ \phi(g_1 U, g_2 U , g_3 U, g_4 U)
    = \phi(g_1,g_2,g_3,g_4)~,~~(U \in G)$$
This condition is equivalent to
\eqn\invariance{  \phi(g_1,g_2,g_3,g_4) = \int  dU
         \phi(g_1U , g_2U , g_3U, g_4U), }
where $dU$ is the invariant Haar measure on $G$
normalized as
$  \int dU =1 $. If $G$ is a finite group, the integral
is replaced by a summation over group elements divided by
the cardinality of the group. With this normalization,
integrals of products of $D^j_{mn}$'s become
\eqn\formula{\eqalign{
   \int  dU \prod_{i=1}^2 D^{j_i}_{m_i n_i}(U)
  & =  \delta^{j_1,j_2}
      g^{j_1}_{m_1 m_2} g^{j_2}_{n_1 n_2} \cr
  \int dU \prod_{i=1}^3 D^{j_i}_{m_in_i}(U)
  & = \left( \matrix{ j_1 & j_2 & j_3 \cr
                      m_1 & m_2 & m_3 \cr } \right)
     \left( \matrix{ j_1 & j_2 & j_3 \cr
                      n_1 & n_2 & n_3 \cr } \right) \cr
  \int dU \prod_{i=1}^4 D^{j_i}_{m_i n_i}(U)
 & = \sum_{j,m,m',n,n'}
       \left( \matrix{ j_1 & j_2 & j \cr
                      m_1 & m_2 & m \cr } \right) g_j^{mm'}
    \left( \matrix{ j & j_3 & j_4 \cr
                      m' & m_3 & m_4 \cr } \right) \cr
  &~~~~~~~~~~~~~~~~~~~~
       \left( \matrix{ j_1 & j_2 & j \cr
                      n_1 & n_2 & n \cr } \right) g_j^{nn'}
    \left( \matrix{ j & j_3 & j_4 \cr
                      n' & n_3 & n_4 \cr } \right). \cr}}
Here we used the Wigner $3j$-symbol which is related
to the Clebsch-Gordan coefficient as
$$ \left( \matrix{ j_1 & j_2 & j_3 \cr
                   m_1 & m_2 & m_3 \cr }  \right)
      = {(-1)^{j_1-j_2-m_3} \over \sqrt{2j_3+1}}
       \langle j_1 j_2 m_1 m_2 | j_3 m_3 \rangle $$
and the metric defined by
$$ \eqalign{
    g^j_{mm'} & = (-1)^{j-m} {1 \over \sqrt{2j+1}} \delta_{m+m',0} \cr
    g_j^{mm'} & = (-1)^{j+m} \sqrt{2j+1} \delta_{m+m',0}. \cr}, $$
By substituting \general\ into \invariance\
and using the above formulae, we see that $\phi$ is expanded as
\eqn\expansion{ \eqalign{
     \phi(g_1,g_2,g_3,g_4) = & \sum
              M^{j_1j_2;j;j_3j_4}_{m_1m_2;m_3m_4}
          D^{j_1}_{m_1n_1}(g_1) \cdots D^{j_4}_{m_4n_4}(g_4) \cr
    &~~~~~ g_{j_1}^{n_1 n_1'} \cdots g_{j_4}^{n_4 n_4'}
         \left( \matrix{ j_1 & j_2 & j \cr
                      n_1' & n_2' & n \cr } \right) g_j^{nn'}
    \left( \matrix{ j & j_3 & j_4 \cr
                      n' & n_3' & n_4' \cr } \right) , \cr} }
where
\eqn\mdef{ M^{j_1 j_2;j;j_3 j_4}_{m_1m_2;m_3m_4}
    = \sum_{n,n',n_i,n_i'} \phi^{j_1...j_4}_{m_1n_1...m_4n_4}
         g_{j_1}^{n_1n_1'} \cdots g_{j_4}^{n_1n_1'}
          \left( \matrix{ j_1 & j_2 & j \cr
                      n_1' & n_2' & n \cr } \right)  g_j^{nn'}
    \left( \matrix{ j & j_3 & j_4 \cr
                      n' & n_3' & n_4' \cr } \right).}

There are several conditions that the coefficient $M$ should obey.
The real-valuedness of $\phi$ implies
$$ \overline{M}^{j_1j_2;j;j_3j_4}_{m_1m_2;m_3m_4}
     = (-1)^{\sum_{i=1}^4(j_i-m_i)}
         M^{j_1 j_2;j;j_3 j_4}_{-m_1-m_2;-m_3-m_4} , $$
where we used the hermiticity of $D^j_{mn}$
$$  \overline{D}^j_{mn}(g) = (-1)^{\sum_{i=1}^4(j_i-m_i)}
                                 D^j_{-m-n}(g). $$
We also require $\phi(g_1,g_2,g_3,g_4)$ to be invariant under
cyclic permutations of any three of its arguments. By the definition \mdef ,
this means that the coefficient satisfies
\eqn\duality{
     M^{j_3 j_1;j;j_2 j_4}_{m_3 m_1;m_2 m_4}
   =  \sum_{j'} (-1)^{j_1+j_2+j_3+j_4}
       \sqrt{(2j+1)(2j'+1)}
     \left\{ \matrix{ j_1 & j_2 & j' \cr
                      j_3 & j_4 & j \cr } \right\}
    M^{j_1j_2;j';j_3j_4}_{m_1m_2;m_3m_4} }
and similar relations obtained by cyclic permutations of
$(j_1,j_2,j_3,j_4)$. Here the symbol $\{ \cdots \}$ in the
right hand side is the Racah-Wigner $6j$-symbol describing
recombination of three angular momenta.

Now let us define our action.
$$ \eqalign{ S={1 \over 2} \int \prod_{i=1}^4  dg_i
                       \phi^2(g_1,g_2,g_3,g_4)
      + {\lambda \over 5!} & \int \prod_{i=1}^{10}  dg_i
                \phi(g_1,g_2,g_3,g_4)
       \phi(g_4,g_5,g_6,g_7) \cr
           &\phi(g_7,g_3,g_8,g_9)
       \phi(g_9,g_6,g_2,g_{10}) \phi(g_{10},g_8,g_5,g_1) \cr} $$
A four-dimensional simplex consists of
five tetrahedra ($3$-simplexes), ten triangles ($2$-simplexes),
ten links ($1$-simplexes) and five vertices ($0$-simplexes).
If we imagine that the function $\phi(g_1,g_2,g_3,g_4)$ is associated
to a tetrahedron with the group elements on its faces,
the kinetic term in the above action is seen as
overlapping of two tetrahedra while the interaction term represents
gluing faces of five tetrahedra to make a $4$-simplex.
Substituting \expansion\ into the above and using \formula ,
the action can be expressed in terms of $M$ as
\eqn\action{ \eqalign{ S(M) &=
     {1\over 2}\sum_{j,m} |M^{j_1j_2,j_3j_4,j}_{m_1m_2,m_3m_4}|^2 \cr
    & -{\lambda \over 5!} \sum_{j,m,n} (-1)^{\sum_{i=1}^{10}(j_i+m_i)}
           (-1)^{\sum_{i=1}^5 (l_i+n_i)}
          \left\{ \matrix{ l_1 & l_2 & l_3 & l_4 & l_5  \cr
                      j_1 & j_2 & j_3 & j_4 & j_5  \cr
                    l_{10}& l_9 & l_8 & l_7 & l_6  \cr } \right\} \cr
           &~~~~M^{j_2 l_2;l_3;l_4 j_3}_{m_2 n_2; -n_4 m_3}
                M^{j_4 l_4,l_5;l_6 j_5}_{m_4 n_4; -n_6 m_5}
                M^{j_2 l_6;l_7;l_8 j_1}_{-m_2 n_6; -n_8 m_1}
                M^{j_3 l_8;l_9;l_{10} j_5}_{-m_3 n_8; -n_{10} -m_5'}
                M^{j_4 l_{10};l_1;l_2 j_1}_{-m_4 n_{10};-n_2 -m_1}. \cr} }
Here
$$ \eqalign{  \left\{ \matrix{ l_1 & l_2 & l_3 & l_4 & l_5  \cr
                      j_1 & j_2 & j_3 & j_4 & j_5  \cr
                    l_{10}& l_9 & l_8 & l_7 & l_6  \cr } \right\}
     & = \sum_{m,n} (-1)^{\sum_{i=1}^{10}(j_i+m_i)}
                         (-1)^{\sum_{i=1}^5 (l_i+n_i)}
          \left( \matrix{ j_1 & l_1 & l_2  \cr
                          m_1 & n_1 & -n_2 \cr } \right) \cr
     & ~~~   \left( \matrix{ j_2 & l_2 & l_3  \cr
                          m_2 & n_2 & -n_3 \cr } \right)
          \left( \matrix{ j_3 & l_3 & l_4  \cr
                          m_3 & n_3 & -n_4 \cr } \right)
          \left( \matrix{ j_4 & l_4 & l_5  \cr
                          m_4 & n_4 & -n_5 \cr } \right) \cr
     & ~~   \left( \matrix{ j_5 & l_5 & l_6  \cr
                          m_5 & n_5 & -n_6 \cr } \right)
          \left( \matrix{ j_2 & l_6 & l_7  \cr
                         -m_2 & n_6 & -n_7 \cr } \right)
          \left( \matrix{ j_1 & l_7 & l_8  \cr
                         -m_1 & n_7 & -n_8 \cr } \right) \cr
     &    \left( \matrix{ j_3 & l_8 & l_9  \cr
                         -m_3 & n_8 & -n_9 \cr } \right)
          \left( \matrix{ j_5 & l_9 & l_{10} \cr
                         -m_5 & n_9 & -n_{10} \cr } \right)
          \left( \matrix{ j_4 & l_{10}& l_1  \cr
                         -m_4 & n_{10}& -n_1 \cr } \right) \cr} $$
is the $15j$-symbol of the third kind \ref\angular{
A.P. Jucys, I.B. Levinson and V.V. Vanagas,
{\it Mathematical Apparatus of the Theory of Angular Momentum}
(Gordon and Breach, 1964);
L.C. Biedenharn, J.D. Louck and P.A. Carruthers,
{\it The Racah-Wigner Algebra in Quantum Theory,
Encyclopedia of Mathematics and its Applications, Volume 9}
(Addison-Wesley, 1981).}. It is not difficult to see that
the $15j$-symbol defined in this way can be expressed
in terms of the $6j$-symbols. Although the
$15j$-symbol does not possess the full rotational
symmetry of the $4$-simplex, the symmetry
is recovered in the interaction term of the action
thanks to the condition \duality\ satisfied by $M$.

A partition function of our model is defined as
\eqn\integral{
    Z = \int \prod_{j,j_i,m_i : \sum m_i=0}
                dM^{j_1j_2;j;j_3j_4}_{m_1m_2;m_3m_4}
                    \exp (-S(M)). }
As in the case of two \two\ and three dimensions \tensor\ \boulatov ,
we expand this integral in powers of $\lambda$. Each term in this
perturbative expansion is constructed from a propagator
derived from the kinetic term in the action and a vertex given
by the $15j$-symbol. The non-vanishing components of
the propagator are
$$ \eqalign{ \langle M^{j_1j_2;j;j_3j_4}_{m_1m_2;m_3m_4}
                \overline{M}^{j_1j_2;j;j_3j_4}_{m_1'm_2';m_3'm_4'}
             \rangle = &
      {1 \over 9} \prod_{i=1}^4
                   \delta^{m_i,m_i'}
                                 \cr
             \langle M^{j_1j_2;j;j_3j_4}_{m_1m_2;m_3m_4}
                \overline{M}^{j_3j_1;j';j_2j_4}_{m_3'm_1';m_2'm_4'}
             \rangle = &
      {1 \over 9} \prod_{i=1}^4
                   \delta_{m_i,m_i'}
        \sqrt{(2j+1)(2j'+1)}
                \left\{ \matrix{ j_1 & j_2 & j' \cr
                                 j_3 & j_4 & j  \cr} \right\} \cr
            \langle M^{j_1j_2;j;j_3j_4}_{m_1m_2;m_3m_4}
                \overline{M}^{j_2j_3;j';j_1j_4}_{m_2'm_3';m_1'm_4'}
             \rangle = &
      {1 \over 9} \prod_{i=1}^4
                   \delta_{m_i,m_i'}
        \sqrt{(2j+1)(2j'+1)}
                \left\{ \matrix{ j_1 & j_2 & j' \cr
                                 j_4 & j_3 & j  \cr} \right\} \cr} $$
and others obtained by the cyclic permutations of $(j_1,j_2,j_3,j_4)$.
By combining these elements together,
the perturbative expansion of the integral \integral\ is expressed as
\eqn\perturb{
  Z = \sum_C {1 \over N_{sym}(C)} \lambda^{N_4(C)} Z_C, }
where the sum $\sum_C$ is over oriented four-dimensional combinatorial
manifolds, $N_{sym}(C)$ is a rank of symmetries of $C$ if any,
$N_4(C)$ is a number of $4$-simplexes in $C$, and $Z_C$ is given
by a summation which can be written symbolically as
 \eqn\summation{
      Z_C =  \sum_{j} \prod_{t:triangles} (2j_t+1)
            \prod_{tetrahedra}
        \{ 6j \} \prod_{4-simplexes} \{ 15j \} .  }

To describe the precise content of the summand in the above,
it is useful to draw a closed $\phi^3$-graph on
each $4$-simplex in $C$ as follows. Each $4$-simplex
has a topology of a four-dimensional ball
and its boundary is covered by five tetrahedra.
On the boundary, we can draw a dual $\phi^4$-diagram
by setting a vertex in each tetrahedron and
and by connecting these vertices by links
dual to the faces of the tetrahedra.
We then split each vertex into two $\phi^3$-vertices
and add an extra-link between them. There are three different ways
to split each $\phi^4$-vertex, and
we make an arbitrary choice of them at this moment.
This defines a closed $\phi^3$-graph on each $4$-simplex.
By attaching a spin $(j,m)$ on each link
and the $3j$-symbol on each vertex of the
$\phi^3$-graph \foot{Although the $3j$-symbol is the cyclic
symmetric, it changes its sign under odd permutation of
the indices. Thus, to be precise, the association of
the $3j$-symbol to the $\phi^3$-vertex is unique modulo
the sign factor. Since each tetrahedron is shared
by two $4$-simplexes, this ambiguity is removed by
choosing the same sign convention on the $\phi^3$-graphes
on the two $4$-simplexes.} and
by summing over the magnetic angular momenta $m$,
we obtain a sum of product of
the $15j$-symbol and several $6j$-symbols \angular .
Different choice of the splittings
results in different combination of the $6j$-symbols.
This procedure is applied to every $4$-simplex in $C$
in such a way that parts of $\phi^3$-graphes of two
neighboring $4$-simplexes match on their common tetrahedron.
As a result, we obtain a sum of product of
$15j$-symbols, each of which is associated to a $4$-simplex in $C$,
and $6j$-symbols which depend on how one splits the $\phi^4$-vertices.
We then perform a sum of this object over
all the spins on the extra-links which were added when we split
the $\phi^4$-vertices.
By using the orthonormality of the $6j$-symbol,
one can show that the result of this summation is
independent of the choice of splitting of the vertices
and depends only on the structure of the complex $C$
and the spins on the links dual to
the $2$-simplexes in $C$. This is what we represent by the product
$\prod \{ 6j \} \prod \{ 15j \}$ in the summand of \summation .

Now we would like to examine the dependence of $Z_C$ on the
simplicial complex $C$. In 1930, Alexander \alex\
defined a set of basic transformations
of simplicial complexes and proved that two
complexes are combinatorially equivalent if
and only if they are connected by a sequence of these transformations.
These basic transformations are called the Alexander moves.
We would like to know how $Z_C$ behaves under the
Alexander moves. Recently Gross and Varsted introduced
another set of transformations which they showed to be
equivalent to the Alexander moves for dimensions less than or
equal to four \move , and we shall use them below.

In four dimensions, there are five basic moves \move\ \four ;

\noindent
(1) Consider a $4$-simplex and add one point at its center.
We then draw five $1$-simplexes which connect this point to the
five vertices of the $4$-simplex.
This decomposes the original $4$-simplex into five $4$-simplexes.

\noindent
(2) Consider two $4$-simplexes sharing one tetrahedron.
There remains a pair of vertices which are not shared by the
$4$-simplexes and we connect them by a $1$-simplex. This
recombines the two $4$-simplexes into three $4$-simplexes.

\noindent
(3) Consider three $4$-simplexes sharing one triangle,
and suppose that there are
three tetrahedron each of which is shared by two of the
$4$-simplexes.
Then there are three $1$-simplexes each of which
belongs to only one of the $4$-simplexes.
The three $4$-simplexes are recombined
into three $4$-simplexes sharing a triangle
spanned by these three $1$-simplexes.

\noindent
The other two moves are obtained by reversing the moves (1) and (2).

To examine the behavior of $Z_C$ under these moves,
it is useful to know that the $15j$-symbols which
appear in \summation\ can be expressed as a sum of
product of the $6j$-symbols \angular .
Then
the Biedenharn-Elliot identity and the orthonormality
of the $6j$-symbols can be used to show that
$Z_C$ is invariant under these moves upto a multiplicative factors.
Rather than showing
the detail of the computations, we present here
another proof, similar to the one used
in \boulatov\ in studying the three-dimensional model.

By using the formulae \formula\ and the definition of
the $15j$-symbol in terms of the $3j$-symbols, we can
rewrite $Z_C$ as follows.
\eqn\gauge{  Z_{C} =  \sum_j \sum_{m,n}
               \prod_{t:triangles} (2j_t+1)
                \prod_{T:tetrahedra}  \int dU_T
               D^{j_{1,T}} (U_T) \cdots
               D^{j_{4,T}} (U_T)  }
where each tetrahedron $T$ carries a group element $U_T$,
and $j_{1,T},...,j_{4,T}$
are the spins on the links of dual to the four faces
of $T$. The matrix elements $D^j$'s are multiplied together
around triangles in $C$. Each triangle $t$
is shared by a finite number of tetrahedra $T_1^{(t)}, T_2^{(t)},...,
T_{n_t}^{(t)}$, and we can perform the sum over the spin-$j_t$
on the link dual to $t$ using the formula
$$  \sum_{j_t} (2j_t+1) {\sl Tr} \left[ D^{j_t}(U_{T_1^{(t)}})
                       \cdots D^{j_t}
                            (U_{T_{n_t}^{(t)}})
               \right]
       = \delta(U_{T_1^{(t)}} \cdots U_{T_{n_t}^{(t)}}, 1). $$
If we regard $\{ U_T \}_{T:tetrahedra}$ as giving
a set of transition functions of a principal
$G$-bundle over the combinatorial manifold, the condition
$U_{T_1^{(t)}} \cdots U_{T_{n_t}^{(t)}}=1$
derived here implies that the
holonomy around the triangle $t$ is trivial.

Now we can examine the behavior of $Z_C$ under the five basic moves of
the simplicial complex. In the first move, the subdivision of
a $4$-simplex into five $4$-simplexes creates ten triangles and
ten tetrahedra.
Correspondingly we introduce ten spin variables $j_1,...,j_{10}$
on the triangles and ten group elements $U_1,...,U_{10}$
on the tetrahedra, which give transition function of
a $G$-bundle in the interior of the original $4$-simplex.
By summing over $j_i$'s, we obtain ten
$\delta$-functions, and they restrict the $G$-bundle in the interior
to be flat. Then the ten group elements can be set to $1$
by redefining the group elements on the original $4$-simplex.
The conditions imposed by these ten $\delta$-functions
are not independent and four of them are redundant,
giving rise to a factor $\delta(1,1)^4$.
Thus $Z_C$ transforms into
$\delta(1,1)^4 Z_C$ under the first move.

If $G$ is a finite
group, $\delta(1,1)$ is simply a number of elements in $G$.
But it is divergent for a continuous group like $SU(2)$.
To regularize this divergence, one may cut off the summation
as $j_i \leq L$ for some large $L$ as in \ponzano .
The divergent factor is removed from the partition function
by multiplying  $\delta(1,1)$ for each $0$-simplex and
dividing by $\delta(1,1)$ for each $1$-simplex (note that
the numbers of the $0$- and $1$-simplexes increases by one and
five in the first move), and we send $L \rightarrow \infty$ after
this.
In this regularization, the invariance under the first move does
not hold at a finite value of $L$ and it is recovered only after taking
the limit $L \rightarrow \infty$. Alternatively
one may use the $q$-deformed version of the model defined below.
When $q$ is a root of unity, $q= e^{2\pi i / (k+2)}$,
the sum over $j$'s is naturally restricted as $j \leq k/2$
and the $q$-analogue of $\delta(1,1)$ is given by
$$ \delta(1,1) = {k+2 \over 2 \sin^2({\pi \over k+2 })}.$$
In the so-called $BF$ model \horowitz\ \blau\
which we will identify as
a continuum limit of our lattice model, this divergent
factor would correspond to a part of its gauge volume.

%
%

In the second move, four triangles are added in the
interior of two $4$-simplexes and summing over spins
on them generates four $\delta$-functions, three of which
are independent. Thus $Z_C$ transforms into $\delta(1,1) Z_C$
under the second move. In this move, the number of the $0$-simplexes
does not change while the $1$-simplexes increase by one.
Thus the divergence is removed in the same way as in the first move.
On the other hand, no redundant
$\delta$-function appears in the third move, and $Z_C$
remains invariant under this move. The fourth and the fifth
moves are the inverses of the first and the second moves,
and $Z_C$ transforms into $\delta(1,1)^{-4} Z_C$
and $\delta(1,1)^{-1} Z_C$ respectively.
Combining these result, the dependence of $Z_C$ on $C$ is
expressed as
\eqn\top{ Z_C = \delta(1,1)^{-N_0(C)+N_1(C)} \chi_G(C), }
where $N_0(C)$ and $N_1(C)$ are numbers of $0$- and $1$-simplexes
in $C$, and $\chi_G(C)$ depends only on the combinatorial
class of $C$.

It is straightforward to define the $q$-deformed version of
the model, by replacing the $15j$-symbols and the $6j$-symbols
by their $q$-analogues. Since the invariance under the Alexander
moves is essentially due to the Biedenharn-Elliot identity and
the orthonormality of the $6j$-symbols, which are also satisfied
by their $q$-analogues \kir , it is likely that
the $q$-analogue model is topological. The detailed analysis
will be reported elsewhere.

The expression \gauge\ suggests that, for a compact group $G$,
$Z_C$ is related to
the partition function of the $BF$ model which is
defined for an oriented manifold $M$ as
$$ Z_{BF} = \int [ d B , d A ] \exp ( i \int_M \langle B,
                          d A + [A , A] \rangle ) ,$$
where $A$ is a connection one-form for a $G$-bundle on
the manifold $M$, $B$ is a ${\cal G}$-valued two-form
(${\cal G}$ is a Lie algebra of $G$) and $\langle , \rangle$
is the invariant bilinear form on ${\cal G}$. If we
integrate over $B$ first, we obtain
\eqn\connection{ Z_{BF} = \int [d A] \delta(dA + [A , A]),}
and $Z_{BF}$ is given by an integral over flat connections.
As we have seen in the above, the set of group elements
$\{ U_T \}_{T:tetrahedra}$ in \gauge\ define a principal
$G$-bundle on the simplicial manifold, and the sum over
$j$'s restricts the $G$-bundle to be flat. Due
to the topological nature of the lattice model, we can
take the triangulation to be infinitely fine.
When $C$ is a smooth triangulation of $M$,
the integral over $U_T$'s in \gauge\  is then
identified with the integral over $A$ in \connection\
in this limit
in the sense of the lattice gauge theory. One may
also regard the spin-$j$ associated to each $2$-simplex
as a discretized version of the two-form $B$.

The action for the $BF$ model has the following
gauge symmetries,
\eqn\symmetry{  \delta A = D \lambda~,~~\delta B = [ B , \lambda  ]
                           + D \omega ,}
where $\lambda$ and $\omega$ are ${\cal G}$-valued
$0$- and $1$-forms, and $D$ is the covariant derivative
with respect to the connection $A$. The Wilson operator
associated to a closed loop $\gamma$ is invariant
under these gauge transformations and thus gives
a physical observable in the theory. One may also
try an operator defined by an integral over
a closed surface $\Sigma$ in $M$ as
$ \exp ( i \int_\Sigma \langle \alpha , B \rangle ) $
for some $\alpha$ in ${\cal G}$. Insertion of this operator
generates a holonomy $e^{i\alpha}$ around the surface $\Sigma$.
This is because the field strength $F$ becomes non-vanishing on $\Sigma$
upon integration over $B$. In the standard Yang-Mills theory,
such an operator has been considered by 't Hooft \thooft ,
so we call it the 't Hooft operator here.
If we take $e^{i\alpha}$ to be in the center of $G$,
the operator becomes invariant under the gauge transformations
\symmetry .

In order to maintain the gauge symmetries
without restricting $e^{i \alpha}$
to be in the center, we consider the following modification to
the 't Hooft operator \foot{An operator associated to a surface
has also been considered in \ref\hortwo{
G.T. Horowitz and  M. Srednicki, {\sl Commun. Math. Phys.} {\bf 130} (1990)
83; I. Oda and S. Yahikozawa, {\sl Phys. Lett.} {\bf B 238}
(1990) 272.}, but it
differs from the one constructed here.}.
Let us choose a normal vector field on $\Sigma$
and displace $\Sigma$ slightly in the direction of the vector field
to obtain another surface $\Sigma'$.
The trajectory swept out by $\Sigma$ then defines
a three-dimensional submanifold $V_\Sigma$ in $M$
bounded by $\Sigma$ and $\Sigma'$.
This procedure is similar to the framing of the Wilson-line in the
three-dimensional Chern-Simons theory \witten .
The integration of the $B$ field in $V_\Sigma$
restricts $A$ to be flat there, and
this induces a flat connection on $\Sigma$ as the boundary value
of $A$. Especially when $\Sigma$
is homeomorphic to $S^2$, the flat gauge field is trivial and
the parallel transport
$\Omega(x,y) = P \exp(i \int_y^x A )$ is defined globally on $\Sigma$.
Thus, in this case, we can define a gauge-invariant operator by
\eqn\surface{ T_\alpha(\Sigma) = \int dU \exp(i \int_\Sigma d^2 x
               \langle B(x) , \Omega(x,y) U \alpha U^{-1} \Omega(y,x)
    \rangle ). }
Because of the $U$-integral, this definition of $T_\alpha(\Sigma)$
is independent of the choice of a point $y$ on $\Sigma$.
It is easy to show that this is invariant under both of the
gauge transformations in the above. In the $BF$ model,
the diffeomorphism invariance is the consequence of these
gauge symmetries. Thus the expectation
value of $T_\alpha(\Sigma)$ is invariant under homotopy
moves of $\Sigma$.
If $\Sigma$ is homeomorphic to a torus or a higher-genus surface,
the flat gauge field is not necessarily trivial and this construction
would need to be modified. In the following, we restrict ourselves to the
case when $\Sigma$ is homeomorphic to $S^2$.

Now let us discuss how to realize $T_\alpha(\Sigma)$ in the lattice model.
In the expression \gauge , each group element $U_T$ is associated
to a tetrahedron $T$ in the simplicial complex $C$,
and each $2$-simplex carries a spin $j_t$.
For each $2$-simplex $t$, the group elements in the
tetrahedra sharing $t$
are multiplied together in the spin-$j_t$ representation $D^{j_t}$
and the trace of the product is taken. Therefore
the holonomy $e^{i \alpha}$
around a given $2$-simplex can be imposed by
inserting $e^{i \alpha}$ in the trace. In the combinatorial manifold,
the surface $\Sigma$ is represented by a collection of $2$-simplexes
$\hat{\Sigma}$. Thus we would like to insert $e^{i \alpha}$
in the traces around all the $2$-simplexes on $\hat{\Sigma}$
in such a way that the topological invariance is maintained.
This can be done, by discretizing the expression \surface ,
as follows.

Let us choose a subcomplex
$\hat{V}_\Sigma$ homeomorphic to $S^2 \times D^2$
($D^2$: two-dimensional disc) in $C$ such that
the surface $\hat{\Sigma} \simeq S^2$ is in the image of
its boundary ($\simeq S^2 \times S^1$). We first
sum over the spin-$j$'s on the $2$-simplexes in the interior
of $\hat{V}_\Sigma$. The $G$-bundle in the interior becomes
flat, and the group elements in the tetrahedra in the interior
can be set to $1$ by redefining the group elements
on the boundary of $\hat{V}_\Sigma$. We then insert $e^{i\alpha}$
on the links dual to the $2$-simplexes on $\hat{\Sigma}$.
Each $2$-simplexes on $\hat{\Sigma}$ is shared by
a pair of tetrahedra in $\partial \hat{V}_\Sigma$, and the
insertion of $e^{i \alpha}$ is made on the links connecting
these pairs.

After inserting $e^{i \alpha}$ on $\hat{\Sigma}$, we integrate
over the group elements in the tetrahedra in $C$.
In the exterior of $\hat{V}_\Sigma$, the integral gives
the product of the $15j$-symbols and the $6j$-symbols
as in \summation . Therefore the product
becomes invariant under the Alexander moves in the exterior
of $\hat{V}_\Sigma$ after summing over spin-$j$'s.
On the other hand, the integral over
the group elements on the boundary $\partial \hat{V}_\Sigma$
generates a product of the $3j$-symbols
corresponding to the $\phi^3$-graph on $\partial \hat{V}_\Sigma$.
The product is summed over the magnetic angular momenta
on the links with
insertions of $e^{i \alpha}$ on the links dual to the $2$-simplexes
on $\hat{\Sigma}$. This defines a function of
spin-$j$'s on the $\phi^3$-graph on $\partial \hat{V}_\Sigma
\simeq S^2 \times S^1$.
By construction, the function depends on the
triangulation on $S^2 \times S^1$, but
its dependence is cancelled by the
contribution from the exterior of $\hat{V}_\Sigma$
after summing over the spin-$j$'s on $\partial \hat{V}_\Sigma$.
The topological invariance is then maintained.

In \viro\ \ooguri , the notion of {\it physical
states} has been
introduced in the topological lattice model
in three dimensions.
In the three-dimensional model,
states on a two-dimensional closed surface are functions of
spin-$j$'s on a $\phi^3$-graph
dual to a triangulation of the surface, and they are called
{\it physical} when they transform
in a definite way specified in \ooguri\ under the Alexander moves
of the triangulation.
It is possible to introduce a similar notion in the four-dimensional
model here.
States in the four-dimensional model
are functions of spin-$j$'s on a $\phi^3$-graph
on a three-dimensional closed manifold, and they
are called physical when they transform in such a way
that the insertion of the state on the three-manifold embedded in
four dimensions does not spoil the topological invariance.
We may regard the 't Hooft operator as
creating a physical state on $S^2 \times S^1$.

For the 't Hooft operator, $S^2 \times S^1$ is the boundary of
$\hat{V}_\Sigma \simeq S^2 \times D^2$, and the $S^1$-part
is contractable in $M$. One may also consider a complementary situation
when the $S^2$-part is contractable while the $S^1$-part may
link with other operators. This gives the Wilson operator
for a closed loop $\gamma$ which is the image of $S^1$.
To be more explicit, we consider a subcomplex $\hat{W}_\gamma$
homeomorphic to $D^3 \times S^1$ ($D^3$: three-dimensional ball)
in $C$ such that the loop $\gamma$ travels through tetrahedra
$T_1$,...,$T_n$ in the interior of $\hat{W}_\gamma$. We then
make a choice of a spin $j$ and insert the discretized version
of the Wilson operator
${\sl tr}[ D^j(U_{T_1}) \cdots D^j(U_{T_n}) ]$ in the integrand
of \gauge . By integrating out the group elements in the
tetrahedra in $\hat{W}_\gamma$ and by
summing over the spins on the links in
the interior of $\hat{W}_\gamma$,
we obtain a function of the spins on the boundary
of $\hat{W}_\gamma$.
It is easy to check that the function defines a physical state
on $S^2 \times S^1$.

We have constructed
the 't Hooft operator associated to a surface $\Sigma$ and the
Wilson operator associated to a loop $\gamma$.
They both create physical states on $S^2 \times S^1$,
so their presence does not spoil the topological invariance
of the lattice model.
Especially their correlation functions are invariant under
ambient isotopy moves of $\Sigma$ and $\gamma$,
and thus give invariants of knots and links.
It should also be possible to define the 't Hooft operator and the
Wilson operator in the $q$-deformed version of the lattice model.
Detailed study of these invariants is left for future investigations.

\vskip 1cm

\centerline{{\bf Acknowledgement}}

I would like to thank T. Eguchi,
K. Higashijima, T. Inami, N. Ishibashi, Y. Yamada, S. Yahikozawa
and T. Yoneya for discussions. I would also like to thank the members of
the theory group in KEK, where part of this work was done, for their
hospitality.

\listrefs

\end